# Impact of Metal ns² Lone Pair on Luminescence Efficiency in Low-Dimensional Halide Perovskites

Hongliang Shi[1], Dan Han[2,3,4], Shiyou Chen[2], and Mao-Hua Du[4*]

[1]Key Laboratory of Micro-Nano Measurement-Manipulation and Physics (Ministry of Education), Department of Physics, Beihang University, Beijing 100191, China

[2]Key Laboratory of Polar Materials and Devices (Ministry of Education) and [3]Department of Physics, East China Normal University, Shanghai 200241, China

[4]Materials Science and Technology Division, Oak Ridge National Laboratory, Oak Ridge, TN 37831, USA

[*]Corresponding Author: Mao-Hua Du (mhdu@ornl.gov)

**Abstract**

Based on first-principles calculations, we show that chemically active metal $ns^2$ lone pairs play an important role in exciton relaxation and dissociation in low-dimensional halide perovskites. We studied excited-state properties of several recently discovered luminescent all-inorganic and hybrid organic-inorganic zero-dimensional (0D) Sn and Pb halides. The results show that, despite the similarity in ground-state electronic structure between Sn and Pb halide perovskites, the chemically more active $Sn^{2+}$ lone pair leads to stronger excited-state structural distortion and larger Stokes shift in Sn halides. The enhanced Stokes shift hinders excitation energy transport, which reduces energy loss to defects and increases the photoluminescence quantum efficiency (PLQE). The presence of the $ns^2$ metal cations in the 0D halide perovskites also promotes the exciton dissociation into electron and hole polarons especially in all-inorganic compounds, in which the coupling between metal-halide clusters is significant.

# I. Introduction

Pb$^{2+}$ and Sn$^{2+}$ based 3D halide perovskites exhibit excellent carrier transport properties.[1-4] CH$_3$NH$_3$PbI$_3$ and related materials have been extensively studied as solar absorber materials.[5-9] The efficient carrier transport in these materials and more generally in ns$^2$-cation-based halides has been attributed partly to the presence of the ns$^2$ lone pair, which leads to dispersive valence and conduction bands and a large static dielectric constant.[10-14] The large dielectric constant provides strong screening of charged defects and impurities, leading to shallow defect and impurity levels that do not affect carrier transport significantly.[15-18] In this paper, we show that the chemically active ns$^2$ lone pair in metal cations also plays an important role in luminescence efficiency in low-dimensional halide perovskites.

Low-dimensional hybrid and all-inorganic halide perovskites have recently attracted intense interests for their novel luminescent properties.[19,20] High PLQE has been reported in many 0D hybrid organic-inorganic halide perovskites,[21-23] owing to the radiative recombination of strongly self-trapped excitons (STEs). For example, (C$_4$N$_2$H$_{14}$X)$_4$SnBr$_6$ exhibits near-unity PLQE.[21] In these 0D hybrid materials, the anionic metal halide octahedra, which are luminescent centers, are separated from each other by large organic cations. As a result, the STE is not only localized but also immobile, suppressing the exciton migration and the energy loss to defects.[24] Compared to hybrid metal halides, the all-inorganic counterparts could be structurally more stable and, thus, more resilient against defect formation under excitation, which is important to the photostability of phosphors and, especially, scintillators under ionizing radiation. Luminescent properties of all-inorganic 0D halide perovskites have been extensively studied with mixed results. The 0D lead bromide Cs$_4$PbBr$_6$ has been studied by many groups for its highly efficient green emission, whose origin, nevertheless, is still under debate.[24-27] The UV emission by STEs in Cs$_4$PbBr$_6$ is, however, weak; suffering from strong thermal quenching at T > 100 K.[28] At room temperature (RT), the STE emission in Cs$_4$PbBr$_6$ is completely quenched. The quenching of the PL in Cs$_4$PbBr$_6$ was attributed to the exciton migration and the subsequent nonradiative recombination at defects, which are abundant in solution-grown halides.[24] On the other hand, related 0D Eu halide perovskites Cs$_4$EuX$_6$ (X = Br, I) were shown to be excellent gamma-ray scintillators with high light yields at RT.[29] In particular, Cs$_4$EuBr$_6$ has the highest light yield under gamma-ray excitation among self-activated scintillators. This should be due to the strong exciton localization at Eu$^{2+}$ ions. More recently, Benin et al. studied luminescent properties in $Cs_4SnX_6$ (X = Br, I),[30] which are isostructural to $Cs_4PbBr_6$. Although there is also thermal quenching in $Cs_4SnBr_6$, it is still highly emissive at RT with PLQE of 15 ± 5%. Increasing the iodine concentration in $Cs_4Sn(Br_{1-x}I_x)_6$ leads to the quenching of the PL at x = 0.5. $Cs_4SnI_6$ is not emissive at RT. As can be seen, the luminescence efficiency in these 0D halide perovskites varies strongly with the chemical composition (especially with the type of the metal ion). Understanding the material chemistry that underpins the different optical properties in 0D metal halides will enable the development of more effective property tuning and optimization methods and the design of new highly-efficient luminescent materials.

In this paper, we performed density functional theory (DFT) calculations with hybrid functionals to study electronic structure as well as exciton and polaron properties in Sn and Pb-based 0D halide perovskites, i.e., $Cs_4SnX_6$ (X = Br, I), $Cs_4PbBr_6$, and $R_4SnBr_6$ (R = $C_4N_2H_{14}Br$). We show that the different PL efficiencies and the different thermal quenching behaviors between 0D Sn and Pb halides lie in the different lone-pair reactivities between $Sn^{2+}$ and $Pb^{2+}$. In comparison to $Cs_4PbBr_6$, the more chemically active $Sn^{2+}$ in $Cs_4SnBr_6$ leads to stronger local structural distortion at the excited state. This gives rise to a larger Stokes shift in $Cs_4SnBr_6$, which in turn limits the exciton migration and energy loss to defects. We further show that the dissociation of excitons into electron and hole polarons in $Cs_4SnX_6$ (X = Br, I), $Cs_4PbBr_6$, and $R_4SnBr_6$ is energetically possible at RT, which is also related to the lone-pair chemistry of $Sn^{2+}$ and $Pb^{2+}$.

## II. Methods

Electronic band structures, density of states (DOS), excitons, and polarons were studied by using hybrid PBE0 functional,[31] which has 25% non-local Fock exchange. Previous PBE0 calculations have shown accurate results in exciton excitation and emission energies in hybrid organic-inorganic halide perovskites.[22,24,32,33] Spin-orbit coupling (SOC) was included in the calculations on $Cs_4PbBr_6$ because it was shown previously that the SOC has a strong effect on Pb-6p-derived conduction band.[34] Following the Franck-Condon principle, the exciton excitation and emission energies were obtained by calculating the total energy differences between the excited and the ground states using PBE0-optimized ground-state and excited-state structures, respectively. The exciton binding energy (relative to a free electron and free hole) was calculated by $\Delta E_b = E(\text{GS}) + E_g - E(\text{exciton})$, where $E(\text{GS})$ and $E(\text{exciton})$ are the total energies of the ground state and the exciton, respectively, and $E_g$ is the band gap.

## III. Results and Discussion

We first compare the electronic band structures of $Cs_4SnBr_6$, $Cs_4SnI_6$, and $Cs_4PbBr_6$, which share the same crystal structure [space group R-3C (#167)]. The Sn compounds have somewhat more dispersive conduction and valence bands than the Pb compound as can be seen in Figure 1. More dispersive bands usually lead to less localized and more mobile excitons and lower PL efficiencies. However, this is not the case here as the exciton emission in $Cs_4SnBr_6$ is more efficient than that in $Cs_4PbBr_6$, which is to be explained below.

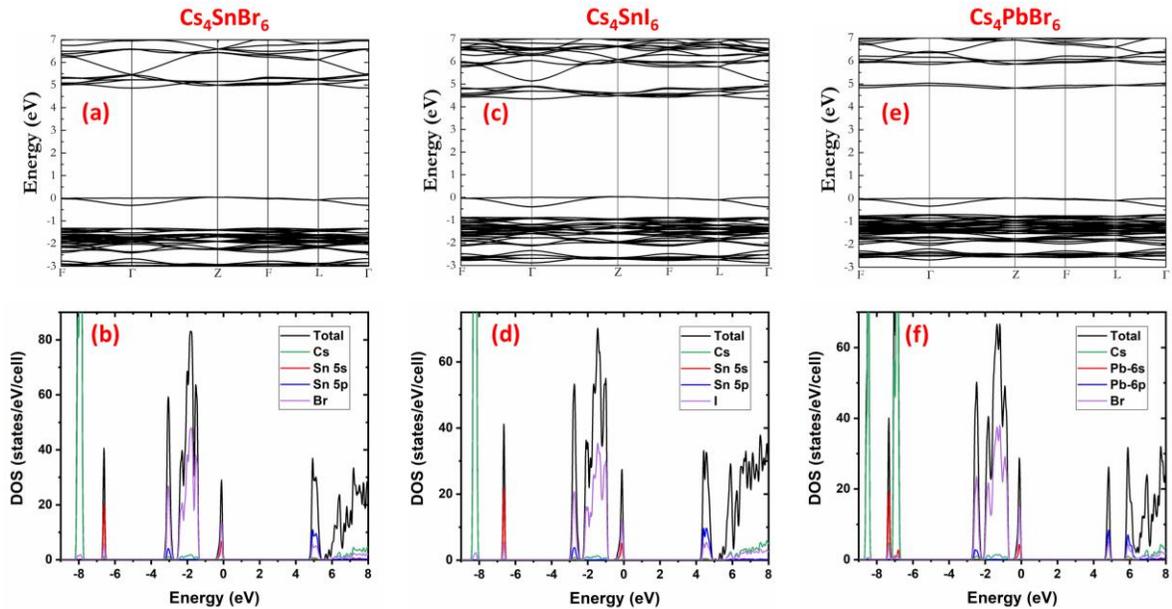

Figure 1. Electronic band structures of (a) $Cs_4SnBr_6$, (b) $Cs_4SnI_6$, and (c) $Cs_4PbBr_6$ obtained by PBE0 calculations. The SOC is included only for (c) $Cs_4PbBr_6$.

The Pb-6p-derived conduction band of $Cs_4PbBr_6$ is split off from other higher-energy Pb-6p bands by the SOC. The conduction band of $Cs_4PbBr_6$ obtained without the SOC resembles that of $Cs_4SnBr_6$. The valence band of these compounds has three components. Taking $Cs_4SnBr_6$ as an example, the band near -3 eV is made up of bonding orbitals between Sn-5p and Br-4p; the top valence band near 0 eV is made up of antibonding orbitals between Sn-5s and Br-4p; the band between the above two is a mixing of largely non-bonding Br-4p orbitals. The energy splitting between the top valence band and the non-bonding halogen-p band indicates the hybridization strength between the metal s lone pair and the halogen p orbitals. This splitting is largest in $Cs_4SnBr_6$ followed by $Cs_4SnI_6$ and $Cs_4PbBr_6$ (Figure 1). Compared to the relatively small energy separation between Sn-5s and Br-4p orbitals, the Sn-5s-I-5p energy separation is larger because the I-5p orbital is higher in energy than the Br-4p orbital; thus, the hybridization between Sn-5s and I-5p in $Cs_4SnI_6$ is weaker compared to that between Sn-5s and Br-4p in $Cs_4SnBr_6$. Turning to $Cs_4PbBr_6$, the Pb-6s orbital is much lower in energy than Sn-5s. As a result, the energy separation between Pb-6s and Br-4p bands in $Cs_4PbBr_6$ is large, resulting in relatively weak hybridization and the smallest energy splitting between the antibonding top valence band and the non-bonding Br-4p band. The strong anti-bonding character of the top valance band in $Cs_4SnBr_6$ has important consequences in the excited-state structural relaxation and optical properties as discussed below.

The calculated valence band width (0.37 eV) in $Cs_4SnBr_6$ is slightly larger than that in $Cs_4PbBr_6$ (0.35 eV). This small difference is consistent with the small difference in the shortest inter-cluster Br-Br distance (4.09 Å vs. 4.07 Å) in $Cs_4SnBr_6$ and $Cs_4PbBr_6$. Thus, the inter-cluster coupling strength at the ground state cannot explain the bright luminescent $Cs_4SnBr_6$ and the non-emissive $Cs_4PbBr_6$ at RT. The valence band width of $Cs_4SnI_6$ is larger at 0.44 eV as a result of better contact between larger-sized I ions.

Figure 2 shows the calculated binding energies of unrelaxed and relaxed excitons (relative to free carriers) in $Cs_4SnX_6$ (X = Br, I), $Cs_4PbBr_6$, and $R_4SnBr_6$. The binding of the unrelaxed exciton in $Cs_4PbBr_6$ (-1.17 eV) and $R_4SnBr_6$ (-1.25 eV) is stronger than that in $Cs_4SnX_6$ (X = Br, I) (-0.92 eV and -0.82 eV); the stronger binding of unrelaxed excitons in the former two compounds is largely the result of narrower conduction and valence bands (see band structures in Figure 1 and Ref. [24]). However, after exciton relaxation, the exciton binding energy of $Cs_4PbBr_6$ (-1.25 eV) becomes significantly weaker than that in $Cs_4SnX_6$ (X = Br, I) and $R_4SnBr_6$ (-1.86 eV, -1.62 eV, and -2.23 eV, respectively), indicating much stronger excited-state structural relaxation in the three Sn halides (Figure 2). The binding of the unrelaxed and relaxed excitons in $R_4SnBr_6$ are both strong, reflecting the fact of both narrow bands and strong exciton relaxation. In these 0D compounds, the exciton localization is strongly favored; the key to the understanding of the different exciton emission efficiencies is the mobility (rather than the localization) of the excitons as discussed below.

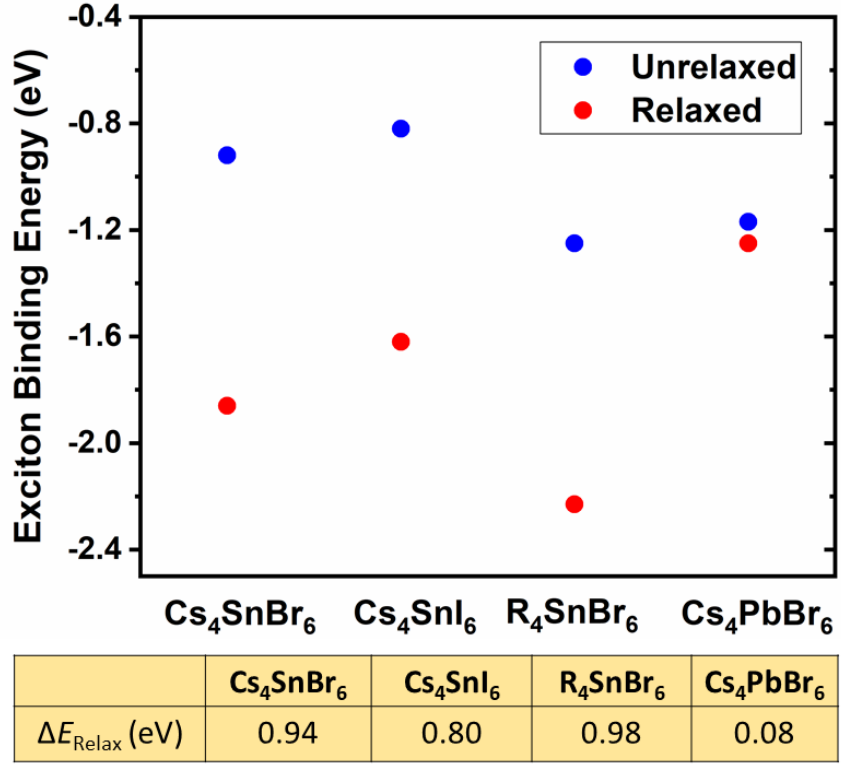

Figure 2. Binding energies of unrelaxed and relaxed excitons relative to free electron and hole as well as the exciton relaxation energy ($\Delta E_{Relax}$), which is the difference between the former two binding energies, in $Cs_4SnBr_6$, $Cs_4SnI_6$, $R_4SnBr_6$, and $Cs_4PbBr_6$. A negative binding energy indicates stable binding.

The calculated exciton excitation/emission energies and Stokes shifts in 0D Sn and Pb halide perovskites are in good agreement with the available experimental values as shown in Figure 3. As a result of the weak exciton relaxation in $Cs_4PbBr_6$, the calculated Stokes shift in $Cs_4PbBr_6$ (0.65 eV) is much smaller than those in the three Sn compounds (1.45 eV – 1.73 eV). A large Stokes shift can reduce the spectral overlap between the excitation and emission; thus, preventing the resonant transfer of excitation energy. [24,35,36] If the resonant condition is not satisfied, the exciton transfer would require phonon assistance, which significantly reduces the energy transport efficiency. The suppression of exciton migration reduces the probability for an exciton to encounter defects; thereby, reducing the nonradiative recombination rate. Increasing temperature broadens the excitation and emission bands. For $Cs_4SnBr_6$ at RT, the tails of the two bands touch each other despite a large measured Stokes shift of 1.35 eV.[30] The measured Stokes shift in $Cs_4PbBr_6$ is only 0.69 eV,[28] which is too small to prevent the spectral overlap as temperature rises. This explains the highly emissive STE in $Cs_4SnBr_6$ [30] as opposed to the absence of STE emission in $Cs_4PbBr_6$ at RT.[28]

The calculated Stokes shift in $Cs_4SnI_6$ (1.45 eV) is smaller than that in $Cs_4SnBr_6$ (1.65 eV). The excitation and emission bands of $Cs_4SnI_6$ are also expected to be broader than those in $Cs_4SnBr_6$ due to the generally softer phonons in iodides than in bromides. Thus, $Cs_4SnI_6$ should suffer from stronger thermal quenching of PL than $Cs_4SnBr_6$, as a result of spectral overlap and resonant excitonic energy transfer. This is consistent with the experimental observation that increasing the iodine concentration in $Cs_4Sn(Br_{1-x}I_x)_6$ leads to the quenching of the RT PL at x = 0.5.

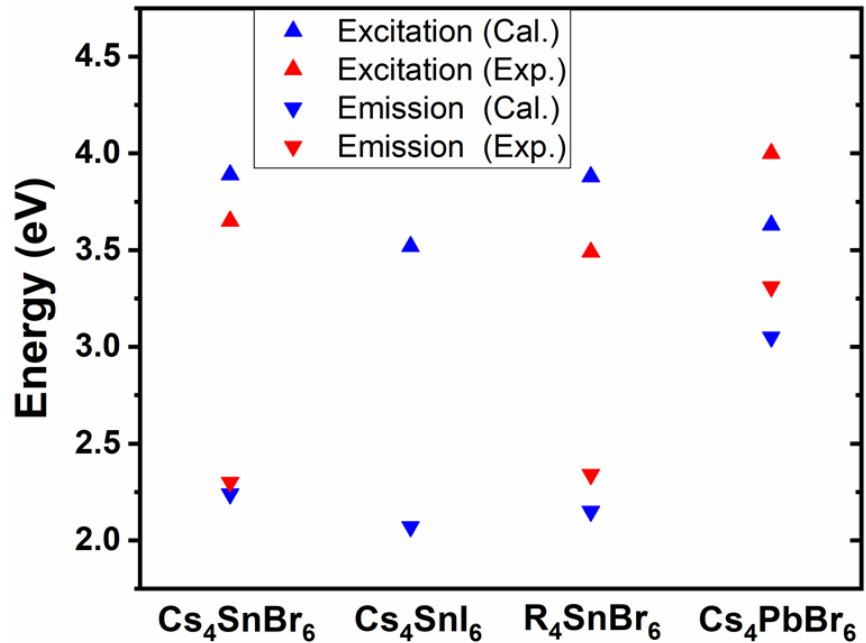

| Stokes Shift (eV) | $Cs_4SnBr_6$ | $Cs_4SnI_6$ | $R_4SnBr_6$ | $Cs_4PbBr_6$ |
|---|---|---|---|---|
| Calculation | 1.65 | 1.45 | 1.73 | 0.58 |
| Experiment | 1.35 | N/A | 1.15 | 0.69 |

Figure 3. Calculated and experimentally measured exciton excitation and emission energies and the Stokes shifts in $Cs_4SnBr_6$, $Cs_4SnI_6$, $R_4SnBr_6$, and $Cs_4PbBr_6$.

As discussed above, a large Stokes shift is important to the PL efficiency. Below we show that the reactivity of the $ns^2$ lone pair has a strong impact on the excited-state structural relaxation and the Stokes shift. We first compare the structures of the STEs in $Cs_4SnBr_6$ and $Cs_4PbBr_6$, which differ only by the type of the $ns^2$ cation. In $Cs_4SnBr_6$, the exciton localization at a $(SnBr_6)^{2-}$ octahedron causes the contraction of the four Sn-Br bonds in the $SnBr_4$ plane (due to the hole localization centered at Sn) and the elongation of the two Sn-Br bonds perpendicular to the plane (due to the electron localization on a Sn-5p orbital). The patterns of the bond length changes in excited states of all four compounds are similar as illustrated in Figure 4. Comparing the STE structures in $Cs_4SnBr_6$ and $Cs_4PbBr_6$, the extent of elongation of the two vertical metal-Br bonds is similar; the two vertical Sn-Br bond length in $Cs_4SnBr_6$ increases from 3.00 Å to 3.52 Å (17.3%) while the two vertical Pb-Br bond length in $Cs_4PbBr_6$ increases from 3.04 Å to 3.52 Å (15.6%). On the other hand, the contraction of the four in-plane Sn-Br bonds in $Cs_4SnBr_6$ [decreasing from 3.00 Å to 2.79 Å (7.0%)] is much stronger than that in $Cs_4PbBr_6$ [decreasing from 3.04 Å to 2.93 Å (3.7%)]. Due to the stronger excited-state structural relaxation in $Cs_4SnBr_6$, the calculated exciton relaxation energy in $Cs_4SnBr_6$ (0.94 eV) is much larger than that in $Cs_4PbBr_6$ (0.08 eV) (Figure 2), which leads to the larger calculated Stokes shift in $Cs_4SnBr_6$ (1.65 eV) than in $Cs_4PbBr_6$ (0.58 eV) (Figure 3).

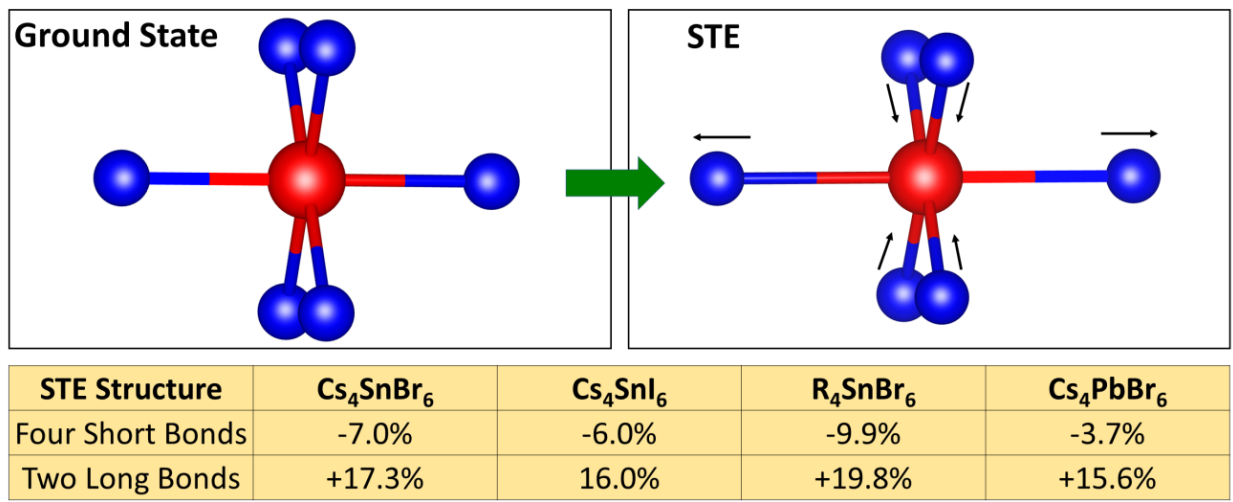

| STE Structure | $Cs_4SnBr_6$ | $Cs_4SnI_6$ | $R_4SnBr_6$ | $Cs_4PbBr_6$ |
|---|---|---|---|---|
| Four Short Bonds | -7.0% | -6.0% | -9.9% | -3.7% |
| Two Long Bonds | +17.3% | 16.0% | +19.8% | +15.6% |

Figure 4. Structure distortion (indicated by black arrows) caused by the STE formation in $Cs_4SnBr_6$, $Cs_4SnI_6$, $R_4SnBr_6$, and $Cs_4PbBr_6$.

The Sn-5s lone pair is chemically more active than the Pb-6s lone pair as the former is closer in energy to the Br-4p band in $Cs_4SnBr_6$ than the latter in $Cs_4PbBr_6$ as can be seen in DOS in Figure 1. As a result, the top valence band of $Cs_4SnBr_6$ has a stronger antibonding character (between the Sn-5s lone pair and Br-4p) than that of $Cs_4PbBr_6$ (between the Pb-6s lone pair and Br-4p). Consequently, creating a hole at the top of the valence band leads to stronger bond contraction in $Cs_4SnBr_6$ than in $Cs_4PbBr_6$. The stronger hole-induced bond contraction in $Cs_4SnBr_6$ further strengthens the hybridization between Sn-5s and Br-4p, resulting in more hole distribution at Sn and stronger Coulomb binding with the electron localized at the Sn-5p orbital. Although $Cs_4SnBr_6$ and $Cs_4PbBr_6$ have the same crystal structure, the stacking of the metal-bromine octahedra in $Cs_4SnBr_6$ is slightly less compact than that in $Cs_4PbBr_6$, which allows more bond length increase of the two vertical bonds upon the addition of an electron to the conduction band. Overall, the excited-state structural relaxation is much stronger in $Cs_4SnBr_6$ than in $Cs_4PbBr_6$, leading to a stronger Stokes shift in the former. This is largely the result of the more chemically active Sn-5s lone pair than the Pb-6s lone pair.

It is shown above that the calculated exciton binding energies relative to free carriers are large for all four 0D metal halides studied here. However, the calculated exciton binding relative to two isolated electron and hole polarons is weak; the binding energies for $Cs_4SnBr_6$, $Cs_4SnI_6$, $R_4SnBr_6$, and $Cs_4PbBr_6$ are 0 eV, -0.19 eV, and -0.09 eV, -0.04, respectively (see Figure 5). Moreover, a polaron pair (an electron and a hole polarons localized on two adjacent metal-halogen octahedra) is more stable than a STE by 0.21 eV ($Cs_4PbBr_6$), 0.17 eV ($Cs_4SnBr_6$), 0.04 eV ($Cs_4SnI_6$), and 0.08 eV ($R_4SnBr_6$). The unusual stability of polarons relative to a STE is also related to the $ns^2$ metal cation. An electron in a STE or in an electron polaron is always localized on a metal-p orbital, causing the elongation of two metal-halogen bonds (as illustrated in Figure 4). However, the hole localization in a STE and in a hole polaron causes different types of structural distortion. In a STE, the hole distributed at the metal-s orbital (which leads to more positively charged metal cation) reduces the lengths of four metal-halogen bonds (Figure 4). On the other hand, when the electron and hole in a STE are separated to two polarons on two metal-halide octahedra, the hole centered at the $ns^2$ cation attract all six halogen anions in the octahedron, leading to six shortened metal-halogen bonds in the hole polaron (rather than four shortened bonds in the case of a STE). This additional Coulomb energy gain compensates the energy loss due to the separation of the electron and the hole. Using $Cs_4SnBr_6$ as an example, the Sn-Br bond length at the ground state is 3.00 Å. In a STE localized at a $SnBr_6$ octahedron, the Sn-Br bond lengths of the two long bonds and the four short bonds are 3.52 Å and 2.79 Å, respectively. In comparison, in an isolated hole polaron, all six Sn-Br bonds are short bonds with the length of 2.79 Å; in an isolated electron polaron, the two long Sn-Br bonds have the bond length of 3.67 Å and the other four bonds have the length of 3.06 Å.

This above analysis reveals that the stability of polarons relative to a STE in 0D halide perovskites containing $ns^2$ cations is related to the fact that the localized electron and hole are both centered at the $ns^2$ cation. If the metal ion in the center of the octahedron is not a $ns^2$ ion such that the hole localization is not centered at the metal cation but rather causes the formation of a halogen-halogen bond (a $V_k$ center),[37] the hole would be localized on the halogen-halogen bond while the electron is localized on the metal cation, which binds the hole. In this case, there is no energy incentive to separate the electron from the hole.

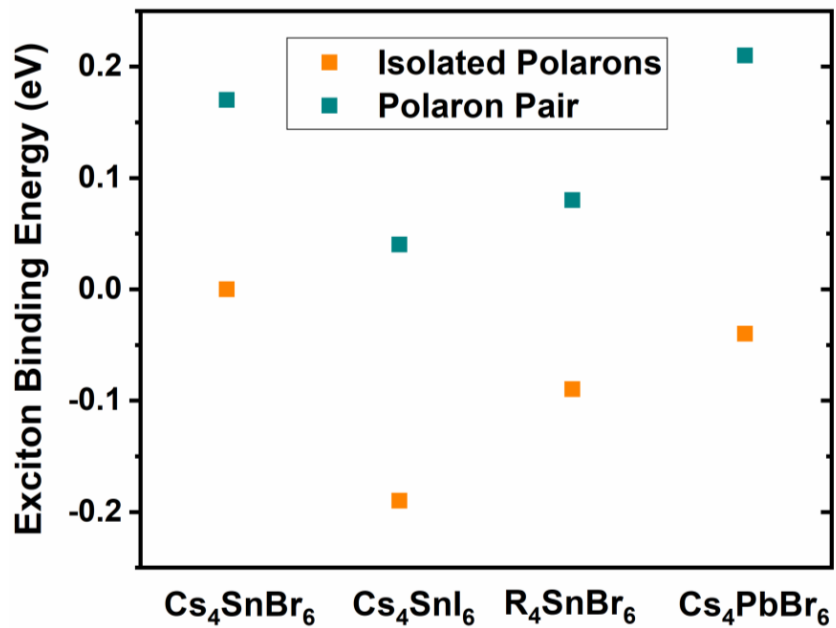

Figure 5. Exciton binding energies relative to isolated electron and hole polarons as well as to electron-hole polaron pairs in $Cs_4SnBr_6$, $Cs_4SnI_6$, $R_4SnBr_6$, and $Cs_4PbBr_6$. A negative binding energy indicates stable binding.

In the above 0D halide perovskites, without structural distortion, an exciton can be localized at one cluster by strong Coulomb binding while the formation of an electron-hole polaron pair is energetically unfavorable. The polaron stabilization must involves strong structural distortion. Thus, optical excitation of an electron-hole pair should lead to an exciton localized on one metal halide cluster rather than a polaron pair on two adjacent clusters. The subsequent excited-state relaxation could lead to a polaron pair if the electron can tunnel to the adjacent cluster, which is subject to a kinetic barrier. The tunneling of an electron polaron from one cluster to another to separate from the hole is more difficult than the exciton tunneling due to the Coulomb binding between the electron and the hole. Even without the Coulomb binding, the diffusion of a charged polaron is in general more difficult and subject to a higher kinetic barrier than the diffusion of a neutral exciton.[38] The polaron tunneling between two adjacent clusters also requires more phonon assistance than the exciton tunneling because, without structure distortion, an electron would prefer delocalization, whereas an exciton still prefers localization due to Coulomb binding. Therefore, increasing temperature is expected to first enable exciton diffusion before the exciton dissociation into polaron pairs becomes possible. We have also calculated the emission energy of a polaron pair in $Cs_4PbBr_6$, which yields 1.79 eV, much lower than the STE emission energy > 3 eV (Figure 3). This shows that the experimentally observed emission is due to the STE localized at one metal-halide cluster rather than the polaron pairs distributed at two adjacent clusters. In all-inorganic $Cs_4PbBr_6$, $Cs_4SnBr_6$, and $Cs_4SnBr_6$, there is relatively strong electronic and vibrational inter-cluster coupling, which may enable the STE dissociation, providing another channel for PL thermal quenching (in addition to the STE diffusion and trapping at defects). In hybrid $R_4SnBr_6$, where the $SnBr_6$ clusters are decoupled by large-sized organic molecules, the STE diffusion and separation are both suppressed; thus, giving rise to a near-unity PLQE.[21,24]

## IV. Summary

This work provides a new understanding on the effect of the metal $ns^2$ lone pair on exciton relaxation and dissociation, which determine the luminescence efficiency, in 0D halide perovskites. We performed hybrid DFT calculations to study exciton emission in 0D Sn and Pb halide perovskites, $Cs_4SnX_6$ (X = Br, I), $Cs_4PbBr_6$, and $R_4SnBr_6$, which contain $ns^2$ metal cations ($Sn^{2+}$ and $Pb^{2+}$). By comparing the $Sn^{2+}$ and $Pb^{2+}$ based compounds, we show that the chemically more active Sn-5s lone pair gives rise to stronger hole-induced excited-state structural distortion and larger Stokes shift. The large Stokes shift in 0D Sn halides hinders the resonant transport of excitation energy; thus, reducing energy loss to defects, which are usually abundant in these solution-grown halides. This explains the generally higher PLQE in 0D Sn halide perovskites than in Pb halides and their thermal quenching behaviors. Our calculations also show that the presence of $ns^2$ metal cations promotes the exciton dissociation into electron and hole polarons in 0D Sn and Pb halide perovskites especially in all-inorganic compounds, in which the coupling between the metal-halide clusters is significant. The separation of the electron and the hole should reduce the efficiency of exciton radiative recombination, providing another mechanism for PL quenching. Our results suggest that the key factor to a high PLQE in 0D halide perovskites is the suppression of exciton and polaron tunneling between inorganic clusters, which are luminescent centers, by a weak inter-cluster coupling and/or a large Stoke shift.

**ACKNOWLEDGMENTS**

The work at ORNL were supported by the U. S. Department of Energy, Office of Science, Basic Energy Sciences, Materials Sciences and Engineering Division. H. Shi was supported by the National Natural Science Foundation of China (NSFC) under Grants No.11604007 and the start-up funding at Beihang University. D. Han and S. Chen were supported by the State Scholarship Fund in China and CC of ECNU.